\documentclass[11pt]{article}
\usepackage{graphicx}
\makeatletter

\@addtoreset{equation}{section}
\newcommand{\be}{\begin{eqnarray}}
\newcommand{\ee}{\end{eqnarray}}
\newcommand{\ba}{\begin{array}}
\newcommand{\ea}{\end{array}}
\newcommand{\nn}{\nonumber}

\makeatletter \@addtoreset{equation}{section} \makeatother

\begin{document}
\vspace{1cm}
\begin{center}
~\\~\\~\\
{\bf  \LARGE Back-Reaction of Classical Fields on \\Black Hole Area  Law }

\vspace{1cm}

                      Wung-Hong Huang\\
                       Department of Physics\\
                       National Cheng Kung University\\
                       Tainan, Taiwan\\
\end{center}
\vspace{1cm}
\begin{center}{\bf  \Large ABSTRACT } \end{center}
(This is the unpublished manuscript while parts are corrected and included in version III of  arXiv:1602.00964  which is submitted to the journal.)\\

We study the back-reaction of classical Maxwell field and massive scalar field  on the BTZ black hole entropy. The exact values of the modification which correct the black hole area law are found.  We discuss the similar properties between the scalar and Maxwell fields which are investigated in both of Coulomb gauge and Lorentz gauge.  The dependences of mass and mode number on the  black hole entropy are illustrated.  We also study the back-reaction  by D branes, which are described by DBI action, and explicitly check that the classical solution which gives  the  black hole entropy precisely corrects the black hole area law.   Our results extend the calculations of  the generalized gravitational entropy proposed in recent by  Lewkowycz  and Maldacena [1].  
\vspace{3cm}
\begin{flushleft}
*E-mail:  whhwung@mail.ncku.edu.tw\\
\end{flushleft}
%%%%%%%%%%%%%%%%%%%%%%%
\newpage
%%%%%%%%%%%%%%%%%%%%%%%
\pagenumbering{arabic}
\tableofcontents
%%%%%%%%%%%%%%%%%%%%%%%
\section{Introduction}
According to the Gibbons and Hawking method [2]  the thermodynamics of black holes  is studied by the Euclidean  partition function with  periodic field $\phi(\tau)= \phi(\tau+\beta)$.  
\be
Z&=&\int D g_{\mu\nu}\int  D \phi~e^{-W_{gravity}(g)-W_{matter}(g,\phi)}\nn\\
&=&\int D g_{\mu\nu}~ e^{-W_{gravity}(g)}~[e^{-W_M(g)}]\nn\\
&=&e^{-[W_g+W_M]}
\ee
The general thermal entropy  is then calculated by [3] 
\be
S_{thermal}&=& -(\beta\partial_\beta-1)~\log Z(\beta)\nn\\
&=& (\beta\partial_\beta-1)(W_g)+(\beta\partial_\beta-1)(W_M)\nn\\
&=& S_g+S_M
\ee
Therefore, the back hole total thermal entropy contains two terms.  The first term is from gravity and called as the Hawking area term.  The second term is the quantum correction part which is from the quantized matter field and can be  called as the entanglement entropy [4-7].  The entanglement entropy constitutes only a (quantum) part of the thermodynamical entropy of the black hole, which is logarithmic divergent [3,8,9]. 

It shall be noticed that in above relation the metric $g_{\mu\nu}$ is a solution to the saddle-point equation 
\be
{\delta (W_g+W_M)\over \delta g_{\mu\nu} }=0
\ee
In the standard investigations the metric $g_{\mu\nu}$ is fixed and $W_M(g)$ could be calculated in many approaches [10,11].  In other word, the back-reaction effect on the spacetime is neglected.  

In the recent paper [1], Lewkowycz  and Maldacena  proposed the  generalized gravitation, in which the back-reaction of classical field on BTZ black  background  is calculated. In [12]  we extended the investigation to the case of Maxwell field.  In this paper we will continue the investigation to the more general solutions on the more realistic  BTZ black hole spacetime which is described by the metric [13,14] 
\be
ds^2=(r^2-\mu)dt^2+ {dr^2\over r^2-\mu}+r^2 d\phi^2
\ee
The associated black hole temperature is  $T=\beta^{-1}$ with
\be
\beta={2\pi\over \sqrt \mu}
\ee
BTZ black hole in above spacetime is called as on coordinate $(t,r,\phi)$. After using the new coordinate
\be
u=\sqrt{r^2-\mu}
\ee
the BTZ geometry can be transformed to
\be
ds^2=u^2dt^2+{du^2\over u^2+\mu}+(u^2+\mu)d\phi^2
\ee
Above spacetime is called as on coordinate $(t,u,\phi)$ which is similar to the metric 
\be
ds^2=r^2d\tau^2+{dr^2\over r^2+n^{-2}}+(r^2+n^{-2})dx^2
\ee
that was used in [1].  Hereafter we replace $\mu$, which is mass of BTZ black hole, by  $m^2$ to save the expression.

In this paper we will consider the classical fields propagating on the above BTZ metric and exactly calculate their contributions on the black hole entropy.  This corresponds to the correction of area law as shown in [1] and derived more detail  in section 2.  In section III we exactly solve the Maxwell field equation on BTZ spacetime and calculate the analytic value of  the entropy.  In section IV we study the case of massive free real scalar field. The dependence of entropy on the field mass is analyzed.  In section V we consider the  DBI action [15] on the BTZ spacetime. We explicitly check that the classical solution which gives  the  black hole entropy precisely corrects the black hole area law.   Last section is devoted to a short summary.  The quantum field correction to the BTZ black hole entropy was studied early in [16,17] and recently in [18] for example.  
%%%%%%%%%%%%%%%%%%%%%%
%%%%%%%%%%%%%%%%%%%%%%
\section{Classical Solution and Area Law}
We first follow the paper [1] (appendix A.3) to prove that the classical solution of Einstein equation  gives  the correction of horizon area in the real black hole spacetime.  While the original proof  in [1] used the coordinate  $(t,u,\phi)$ in Eq(1.7) we will in here adopt the black hole coordinate  $(t,r,\phi)$.  As the metric in the coordinate Eq(1.4) explicity shows the black mass $\mu$ we can easily see the temperature  in our proof. \\

Consider the matter field Lagrangian   $ {\cal L}(g_{\mu \nu},\Phi,\nabla_{\mu} \Phi) $ on  Euclidean spacetime with $t=t+\beta$.  Its thermal entropy  (It is called as generalized gravitational entropy in [1])  can be calculated by
\be
 S_T&=& -\beta \partial_\beta \Big[ \log Z^{\rm matter}\Big]+ \log Z^{\rm matter}\nn\\
& = & -\beta \partial_\beta \Big[ \int dt \int d\vec x \sqrt{g} {\cal L}_{\rm matter}\Big] +\Big( \int dt \int d\vec x  \sqrt{g} {\cal L}_{\rm matter}\Big)\nn\\
& = & -\beta^2 \partial_\beta \Big[\int d\vec x  \sqrt{g} {\cal L}_{\rm matter}\Big]\nn\\
& = & -\beta^2 \Big[\int d\vec x  \Big({\partial \sqrt{g} {\cal L}_{\rm matter} \over \partial g^{\mu \nu}}\frac{\partial g^{\mu \nu}}{\partial \beta}+ {\delta \sqrt{g} {\cal L}_{\rm matter} \over \delta \Phi} \delta (\partial_\beta \Phi)\Big)\nn\nn\\
&=&-{\beta\over 16\pi G}~ \int dt~d\vec x  \sqrt{g}~G_{\mu\nu} { \partial g^{\mu\nu } \over \partial \beta }
\ee
in which we have used matte field equation ${\delta \sqrt{g} {\cal L}_{\rm matter} \over \delta \Phi}=0$ and Einstein field equation $G_{\mu \nu}=8\pi GT_{\mu \nu}$.
\\

 To proceed, we also follow [1] to prove that  above result is just the area of the horizon. 
\\

 As we consider the system on fixed metric, The probe classical matter scalar field is small, say $\phi\sim~\eta$,  then the metric will get correction $O(\eta^2)$. Thus the  gravitational part of the action has not term of order $\eta^2$ and  $\partial_\beta$ derivative of the gravitational part vanishes at order $\eta^2$. Thus
\be
 \partial_\beta [ \log Z^{\rm Grav}]&=&0= \int dt~d\vec x~\sqrt{g} G_{\mu\nu} {\partial g^{\mu\nu } \over \partial \beta }  - \int dt~d\vec y~  \sqrt{g} \nabla_\mu \partial_\beta g^{\mu r}
\ee
in which $d\vec y$ is $d\vec x$ while removes the radius coordinate $dr$. Thus
\be
 \int dt~d\vec x~\sqrt{g} G_{\mu\nu} {\partial g^{\mu\nu } \over \partial \beta }&=& \int dt~d\vec y~ \sqrt{g} \nabla_\mu \partial_\beta g^{\mu r}|_{r=r_H}
\ee
The right integration gives the surface integration, i.e, the area of the horizon, ${\cal A_H}$, or more precisely the area of the horizon at order $\eta^2$. To see this property, let us consider the following black hole spacetime with horizon radius $r_H$ at which $C(r_H)=0$
\be
ds^2=-C(r)dt^2+{dr^2\over C(r)}+\cdot\cdot\cdot
\ee
in which ``$\cdot\cdot\cdot$" does not contain coordinate $dt$ and $dr$.  Above black hole has inverse temperature $\beta$ with 
\be
C'(r_H) =4\pi \beta^{-1}
\ee
and
\be
 \int dt~d\vec y~ \sqrt{g} \nabla_\mu \partial_\beta g^{\mu r}|_{r=r_H}&=& \int dt~d\vec y~  \sqrt{g}~ \partial_\beta  (4\pi\beta^{-1})\nn\\
&=&-4\pi \beta^{-1}~\int d\vec y~  \sqrt{g_{\vec y}}\nn\\
&=&-4\pi \beta^{-1}~{\cal A_H}
\ee
Comparing to the original paper [1]  our general formula shows the explicity factor $4\pi\beta^{-1}$. Therefore
\be
S_T={1\over 4G}{\cal A_H}
\ee
 These complete our proof. 
%%%%%%%%%%%%%%%%%%%%
\section {Maxwell Field on BTZ Spacetime }
In this section we first investigate the Maxwell  field in Coulomb gauge on the BTZ spacetime of coordinates $(t,r,\phi)$ or $(t,u,\phi)$. Next, we investigate the Maxwell  field in Lorentz gauge on the BTZ spacetime of coordinates $(t,r,\phi)$ or $(t,u,\phi)$.
\subsection {Maxwell Field in Coulomb Gauge}
\subsubsection {Coulomb Gauge on Coordinates $(t,r,\phi)$}
We first consider BTZ black hole on coordinate $(t,r,\phi)$. The conventional action of Maxwell field is
\be
A=-{1\over 4}\int d^3x\sqrt g~F_{\mu\nu}F_{\lambda\delta}g^{\mu\lambda}g^{\nu\delta}
\ee
We choose the gauge of $A_\phi(r)=A_r(r)=0$ \footnote {In fact, if choosing $A_\mu = \{0,0,\cos( n \phi)  A_\phi (r)\}$ then it give the general solution $ A_\phi (r)=C_1+C_2\log(r^2-m^2)$  which become divergent on horizon. Also, $A_r$ has not dynamics because that $F_{\mu\nu}$ is an antisymmetric tensor  and there is not $A'_r$ term} and assume that the Maxwell field  is 
\be
A_\mu = (\cos( n \phi)  A_t (r),0,0)
\ee
Then the action becomes
\be
A&=&-{\pi\beta\over 8}\int dr\Big[r A_t'(r)^2+{n ^2 A_t(r)^2\over r(-m^2+r^2)}\Big]\nn\\
&=&-{\pi\beta\over 8}\int dr~\Big[r A_t(r)A'_t(r)\Big]'
+{\pi\beta\over 8}\int dr~\Big[A_t(r)(r A_t(r))'-{n^2 A_t(r)^2\over r(-m^2+r^2)}\Big]\nn\\
\ee
The first bracket is the surface term and will contribute to the on-shell gravitational action, which is considered later.  After the variation the second bracket gives the  field equation 
\be
rA''_t(r)+A'_t(r)+{n^2 A_t(r)\over -m^2 r + r^3}=0
\ee
The field equation has following two solutions 
\be
A^{(1)}_t(r)&=&r^{\frac{n}{m}} \, _2F_1\left(\frac{n }{2
 m},\frac{n }{2 m};1+\frac{n}{m};\frac{r^2}{m^2}\right)\\
A^{(2)}_t(r)&=&r^{-\frac{n }{m}} \,
   _2F_1\left(-\frac{n }{2 m},-\frac{n }{2
   m};1-\frac{n }{m};\frac{r^2}{m^2}\right)
\ee
While the second solution which being divergent on horizon is unphysical we adopt the first solution which is zero on horizon.  After normalizing the first solution by $A^{(1)}_t(\infty)=\log(r)$, like as in [1], we use the following normalized function to calculate its correction to the BTZ entropy.
\be
A_t(r)&=&({-1\over m^2})^{\frac{n }{2 m}} r^{\frac{n }{ m}}
  \Gamma\Big[\frac{n }{2 m}\Big] \Gamma\Big[1 + \frac{n }{2 m}\Big]  \,
   _2F_1\Big(\frac{n }{2 m}, \frac{n }{2 m}, 1 + \frac{n }{m}, \frac{r^2 }{m^2}\Big){1\over 2 \Gamma\Big[{m + n\over m}\Big]}
\ee
Substituting the solution into the on-shell gravitational action we find 
\be
\log Z(\beta)&=&-{\pi\beta\over 8}\int dr~\Big[r A_t(r)A'_t(r)\Big]'=-{\pi\beta\over 8}~[r A_t(r)A'_t(r)\Big]_{r\rightarrow \infty}\nn\\
&=&-{\pi\beta\over 8}~\frac{-n  \log \left(m^2\right)-2 (m-n \log (r))-2n \left(\psi ^{(0)}\left(\frac{n}{2 m}\right)+\gamma\right)}{2 n}
\ee
The terms linear in $\beta$ including divergent terms should be subtracted [1].  However, they do not contribute to the entropy. The associated black hole entropy therefore becomes 
\be
S_{A}^{(I)}(m,n,\beta)&=&-\partial_\beta\log Z(\beta)+\log Z(\beta)\nn\\
&=&\frac{\pi ^2 \left(2 m (m+n)-n^2 \psi^{(1)}\left(\frac{n}{2 m}\right)\right)}{m^2 n}
\ee
Let us discuss  some properties about  $S_{A}^{(I)}(m,n,\beta)$ :\\

(1) In the limit of  $m=n=1$we find that 
\be
S_{A}^{(I)}(1,1,\beta)=4\pi^2-{\pi^4\over 2}
\ee
which is just that calculated in [12] and exactly matches with the value from scalar field calculated in the original paper [1].

(2) Notice that  $S_{A}^{(I)}(m,n,\beta)$ is negative.  The property was first found in [1].  

(3) The expansion of  $S_{A}^{(I)}(m,n,\beta)$ for  small or large mass is 
\be
S_{A}^{(I)}(m,n,\beta)&=&-{2 \pi^2 m\over 3 n^2} + {\cal O}(m^2)\\
S_{A}^{(I)}(m,n,\beta)&=&-{ \pi^2 \over  n} + { \pi^2 \over  m}+ {\cal O}(1/m^2)
\ee 
Thus, the entropy value will approach to $-\pi^2/n$ for large black hole mass.  For clear we plot figure 1 to show how the Maxwell entropy depends on mass.
\\
\\
\\
\scalebox{1}{\hspace{3cm}\includegraphics{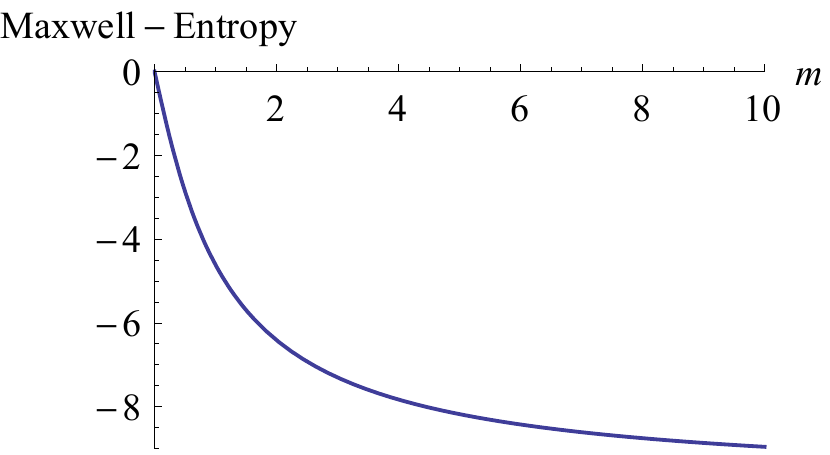}}
\\          
{\it Figure 1:  Dependence of Maxwell entropy on $m$  ($n=1$).}
\\

(4) The expansion of  $S_{A}^{(I)}(m,n,\beta)$ for  small or large mode number, which is an integral number, is 
\be
S_{A}^{(I)}(m,n,\beta)&=&-{ \pi^2 \over  n} + { \pi^2 \over  m}+ {\cal O}(n^2)\\
S_{A}^{(I)}(m,n,\beta)&=&-{2 \pi^2 m\over 3 n^2} + {\cal O}(1/n^3)
\ee 
Thus, the entropy value will approach to $-{2 \pi^2 m\over 3 n^2}$ for large mode number $n$. For clear we plot  figure 2  to show how the Maxwell entropy depends on  mode number $n$.
\\
\\
\\
\scalebox{1}{\hspace{3cm}\includegraphics{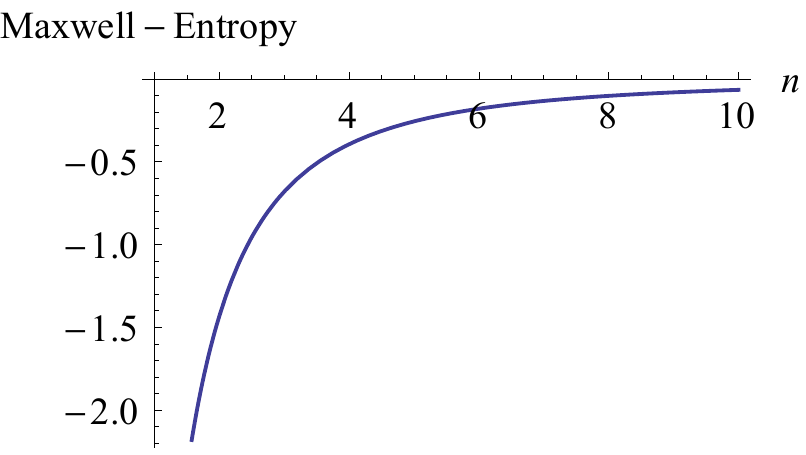}}
\\        
{\it Figure 2: Dependence of Maxwell entropy on $n$  ($m=1$).}
\\

(5) It is interesting to see  that the expansion of small (large) mass is equal to the expansion of  large (small) mode number.  This is because that $n S_{A}^{(I)}(m,n,\beta)$ is a function of $ n/m$.

(6) Above results show that $n$  can not be zero.  To find the solution of $n=0$ mode (called as zero mode) we turn to the possible form
\be
A_\mu = (A_t(r),0,0)
\ee
The field equation is
\be
r A_t''(r)+A_t'(r)=0
\ee
which has exact solution
\be
A_t(r)=C_1+C_2~\log(r)
\ee
The conditions of  $A_\mu=0$ on horizon and $A_t(\infty)=\log(r)$  lead to the normalized solution
\be
A_t(u)=\log( r)-\log(m)
\ee
Substituting the solution into the on-shell gravitational action we find 
\be
\log Z(\beta)&=&{\pi\beta\over 2}\Big[\log( r)-\log(m)\Big]
\ee
Therefore the black hole entropy  is 
\be
S_{A}^{(I)}(m,0,\beta)&=&{ \pi^2 \over m}
\ee 
Not that the correction entropy form zero mode of Maxwell field is positive and is a decreasing function with respective to $m$.

We next  consider BTZ black hole on coordinate $(t,u,\phi)$.
%%%%%%%%%%%%%%%%%%%%%%%
\subsubsection {Coulomb Gauge on Coordinates $(t,u,\phi)$}
On coordinate $(t,u,\phi)$ if we choose the gauge of $A_\phi=A_r=0$ and  assume that the Maxwell field  is $A_\mu = (\cos( n \phi)  A_t(u),0,0)$, then we could not find the exact solution.   Thus, we turn to choose the gauge of $A_t=A_r=0$ and  assume that the Maxwell field  is 
\be
A_\mu = (0,0,\cos(\omega t)  A_\phi(u))
\ee
Then the action becomes
\be
A&=&-{\pi\beta\over 8}\int du\Big[u A_t'(u)^2+{\omega ^2 A_t(u)^2\over u(m^2+u^2)}\Big]\nn\\
&=&-{\pi\beta\over 8}\int du~\Big[u A_t(u)A'_t(u)\Big]'
+{\pi\beta\over 8}\int du~\Big[A_t(u)(u A_t(u))'-{\omega^2 A_t(u)^2\over u(m^2+u^2)}\Big]\nn\\
\ee
The first bracket is the surface term and will contribute to the on-shell gravitational action, which is considered later.  After the variation the second bracket gives the  field equation 
\be
uA''_t(u)+A'_t(u)-{\omega^2 A_t(u)\over m^2 u + u^3}=0
\ee
The field equation has following two solutions 
\be
A^{(1)}_t(u)&=&u^{\frac{\omega}{m}} \, _2F_1\left(\frac{\omega}{2
 m},\frac{\omega }{2 m};1+\frac{\omega}{m};-\frac{u^2}{m^2}\right)\\
A^{(2)}_t(u)&=&u^{-\frac{\omega}{m}} \,
   _2F_1\left(-\frac{\omega}{2 m},-\frac{\omega}{2
   m};1-\frac{\omega}{m};-\frac{u^2}{m^2}\right)
\ee
While the second solution which being divergent on horizon is unphysical we adopt the first solution which is zero on horizon.  After normalizing the first solution by $A^{(1)}_t(\infty)=\log(u)$, like as in [1], we use the following function to calculate its correction to the BTZ entropy.
\be
A_t(u)&=&-\frac{\sqrt{\pi } 2^{-\frac{\omega}{m}-1} m^{-\frac{\omega }{m}}u^{\frac{\omega}{m}} \Gamma \left(\frac{\omega }{2 m}\right) \,
   _2F_1\left(\frac{\omega}{2 m},\frac{\omega }{2 m};\frac{\omega
   }{m}+1;-\frac{u^2}{m^2}\right)}{\Gamma \left(\frac{m+\omega}{2
   m}\right)}
\ee
Substituting the solution into the on-shell gravitational action we find 
\be
\log Z(\beta)&=&-{\pi\beta\over 8}~\frac{-\omega  \log \left(m^2\right)-2 (m-\omega \log (u))-2\omega  \left(\psi ^{(0)}\left(\frac{\omega }{2 m}\right)+\gamma\right)}{2\omega }
\ee
which is exactly the value in coordinate $(t,r,\phi)$ after $u\rightarrow r$ and it gives the same black hole entropy i.e.
\be
S_{A}^{(II)}(m,\omega,\beta)&=&-\partial_\beta\log Z(\beta)+\log Z(\beta)\nn\\
&=&\frac{\pi ^2 \left(2 m (m+\omega)-\omega^2 \psi^{(1)}\left(\frac{\omega}{2 m}\right)\right)}{m^2 \omega}
\ee 

Like as that in previous case $S_{A}^{(I)}(m,\omega,\beta)$,  $S_{A}^{(II)}(m,n,\beta)$ becomes singular if  $\omega =0$.  To find the solution of  zero mode we try  the possible form
\be
A_\mu = (0,0,A_\phi(r))
\ee
It has general solution
\be
A_t(r)=C_1+C_2\log(u)
\ee
which, however, becomes divergent on horizon $u=0$.  Thus, let us turn to another possible solution
\be
A_\mu = (A_t(u),0,0)
\ee
The field equation becomes
\be
uA_t''(u)+A_t'(u)=0
\ee
and normalized solution is 
\be
A_t(u)={1\over 2} \log(u^2+m^2)-{1\over 2} \log(m^2)
\ee
Substituting the solution into the on-shell gravitational action we find 
\be
\log Z(\beta)&=& {\pi \beta\over 4} \log(u^2+m^2)-{\pi \beta\over 4} \log(m^2)
\ee
Therefore the black hole entropy  is 
\be
S_{A}^{(II)}(m,0,\beta)&=&{ \pi^2 \over m}
\ee 
The value is same as the zero mode on the coordinate  $(t,r,\phi)$.  
%%%%%%%%%%%%%%%%%%%%%
%%%%%%%%%%%%%%%%%%%%%%
\subsection {Maxwell Field in Lorentz Gauge}
In this section we study the Maxwell field on BTZ spacetime in the Lorentz gauge and compare it with that on Coulomb gauge.  We consider the action 
\be
A_{F^2}&=&- {1\over 4}\int d^3x {\sqrt g}~\Big[(\nabla_\mu A_\nu-\nabla_\nu A_\mu)(\nabla^\mu A^\nu-\nabla^\nu A^\mu)\Big]\nn\\
&=&- {1\over 2}\int d^3x {\sqrt g}~\Big( \nabla_\mu \Big[A_\nu(\nabla^\mu A^\nu-\nabla^\nu A^\mu)\Big] + {1\over 2}\int d^3x {\sqrt g}~\Big(A^\nu \nabla_\mu \Big[\nabla^\mu A^\nu-\nabla^\nu A^\mu\Big]\Big)\nn\\
&=&- {1\over 2}\int d^3x {\sqrt g}~\Big( \nabla_\mu \Big[A_\nu(\nabla^\mu A^\nu-\nabla^\nu A^\mu)\Big] + {1\over 2}\int d^3x {\sqrt g}~A_\mu \Big(\nabla^2~\delta^\mu_\nu- R^{\mu}_{\nu}-\nabla^\mu \nabla_\nu\Big)A^\nu\nn\\
\ee
in which $\nabla_\mu A_\nu \equiv \partial_\mu A_\nu -\Gamma_{\mu\nu}^\lambda A_\lambda$ and $[D_\mu, D_\nu] A_\lambda=R_{\mu\nu\lambda}^c A_c$.  The first bracket is the surface term and will contribute to the on-shell gravitational action, which is considered later.  After the variation the second bracket gives the  field equation.  

Therefore, in the Lorentz gauge, $\nabla_\nu A^\nu=0$, the field equation becomes
\be
(\nabla^2~\delta^\mu_\nu- R^{\mu}_{\nu})A^\nu=0
\ee
in which $R^{\mu}_{\nu}={\rm Diagonal}(-2,-2,-2)$.  

To proceed, we can consider the following possible non-zero mode solution in coordinate $(t,t,\phi)$  (The solution in coordinate $(t,u,\phi)$ has same property)
\be
A_\mu =(\cos(n\phi)  A_t (r),0,0)
\ee
which automatically satisfies the Lorentz gauge.   In this case the field equation of $A_t(r)$ is exactly that in Coulomb gauge. Substituting  the solution into the above on-shell gravitational action we get the same entropy as that in Coulomb gauge.

Next,  we can, as before, consider the possible zero-mode solution of
\be
A_\mu =(A_t(r),0,0)
\ee
which automatically satisfies the Lorentz gauge and has a solution  as that in Coulomb gauge.  Through the same analysis we find that the zero-mode contributes the same entropy as that in Coulomb gauge.  These conclude that the generalized black hole entropy calculated in Coulomb gauge is same as that calculated in Lorentz gauge.
%%%%%%%%%%%%%%%%%%%%%%%
%%%%%%%%%%%%%%%%%%%%%%%
%%%%%%%%%%%%%%%%%%%%%%%
\section {Massive Real Scalar Field on BTZ spacetime }
The action of massive real scalar field is
\be
A^{\Phi}&=&{1\over 2}\int dx^3 \sqrt g~g^{ab}\Big[\partial_a\Phi \partial_b\Phi+M^2\Phi^2\Big]\nn\\
&=&{1\over 2}\int dx^3 \partial_a[\sqrt g~g^{ab}\Phi \partial_b\Phi]-{1\over 2}\int dx^3~ \Phi\Big[\partial_a(\sqrt g~g^{ab} \partial_b\Phi-M^2\Phi)\Big]
\ee
The first bracket is the surface term and will contribute to the on-shell gravitational action, which is considered later.  After the variation the second bracket gives the scalar field equation : $\nabla^2 \Phi-M^2\Phi=0$.  In this section we will work only in coordinate $(t,u,\phi)$ because that we could not find the general analytical property in coordinate $(t,r,\phi)$. 
%%%%%%%%%%%%%%%%%%%%%%%
%%%%%%%%%%%%%%%%%%%%%%%
\subsection {Higher-Mode of Massive Real Scalar Field }
We adopt the  general assumption about the real scalar field 
\be
\Phi= \cos (\omega t) f(u)
\ee 
In this case the field equation becomes
\be
 u^2 (m + u^2) f''(u) +  u(m + 3 u^2) f'(u) -(\omega^2+M^2u^2) f(u) =0
\ee
The field equation has following two solutions 
\be
 f_1(u)&=&u^{\omega\over m}  \, _2F_1\Big( {1\over 2} \Big(2+{\omega\over m}- {\Delta\over 2}\Big), {\Delta\over 2} +{\omega\over 2 m}, {m + \omega\over    m}, -{u^2\over m^2}\Big)\\
f_2(u)&=& u^{-\omega\over m}\, _2F_1\Big(-{\omega+m(\Delta-2)\over 2 m},  {\Delta\over 2} - {\omega\over 2 m}, 1 - {\omega\over m}, -{u^2\over m^2}\Big)
\ee
where $\Delta \equiv 1+\sqrt{1 + M^2}$. While the second solution which being divergent on horizon is unphysical we adopt the first solution which is finite  on horizon.  

 To proceed, we normalize the first solution like as in [1] and  use the following function to calculate its correction to the BTZ entropy.
\be
f(u)={\Gamma[ {\omega\over 2 m}+ {\Delta\over 2 }]^2\over \Gamma [1+\frac{\omega}{m}]\Gamma [\Delta -1]}~(m)^{-2+\Delta-{\omega\over m}}~(u)^{\omega\over m}\, _2F_1\Big( {1\over 2} \Big(2+{\omega\over m}- {\Delta\over 2}\Big), {\Delta\over 2} +{\omega\over 2 m}, {m + \omega\over    m}, -{u^2\over m^2}\Big)
\ee
Substituting the solution into the on-shell gravitational action we find 
\be
\log Z(\beta)&=&{1\over 2}\int dx^3 \partial_a\Big[\sqrt g~g^{ab}\Phi \partial_b\Phi\Big]\nn\\
&=&{1\over 2}\int d^3  \Big[\partial_t(\sqrt g~g^{tt}\Phi \partial_t\Phi)+\partial_r(\sqrt g~g^{rr}\Phi \partial_r\Phi)+\partial_\phi(\sqrt g~g^{\phi\phi}\Phi \partial_\phi\Phi)\Big]\nn\\
&=&{1\over 2}\int dt d\phi \Big[\sqrt g~g^{uu}\Phi \partial_b\Phi\Big]_{u\rightarrow\infty}={\pi\over 2}\beta~\Big[u~(u^2+m)\Phi \partial_b\Phi\Big]_{u\rightarrow\infty}
\ee
After lengthy calculations the associated black hole entropy  becomes 
\be
&&S_{\Phi}(M,m,\omega,\beta)=-\partial_\beta\log Z(\beta)+\log Z(\beta)\nn\\
&&=\frac{4 \pi ^3 (\Delta -1) \csc (\pi  \Delta ) m^{2 \Delta -4} \Big(-\omega
   H[{\frac{1}{2} \Big(\frac{\omega}{m}+\Delta -2\Big)}]+\omega H[{\frac{\omega-m
   \Delta }{2 m}}]+2 (\Delta -1) m\Big) \Gamma \Big(\frac{\omega+m \Delta }{2
   m}\Big)^2}{\Gamma (\Delta )^2 \Gamma \Big(\frac{1}{2}
   \Big(\frac{\omega}{m}-\Delta +2\Big)\Big)^2}\nn\\
\ee
where $H[X]$ is the Harmonic number at  $X$.  Now, let us discuss the properties of  $S_{\Phi}(M,m,\omega,\beta)$.

(1) In the case of  $m=\omega=1$ then
\be
S_{\Phi}(M,1,1,\beta)=\frac{4 \pi^3 \csc (\pi  \Delta ) \Big(2 (\Delta -2) \Delta -\pi  (\Delta -1) \tan  \Big(\frac{\pi  \Delta }{2}\Big)\Big) \Gamma    \Big(\frac{\Delta +1}{2}\Big)^2}{\Gamma\Big(\frac{3}{2}-\frac{\Delta }{2}\Big)^2 \Big(\Gamma (\Delta )\Big)^2}
\ee
In the case of  $M=0$,  $S_{\Phi}(0,1,1,\beta)$  is just the value in Maxwell field.  This is because that on 3D the maxwell field is dual to scalar field.

(2) The entropy will be divergent when $\Delta =4,6...$, as shown in figure 3.
\\
\\
\\
\scalebox{1}{\hspace{3cm}\includegraphics{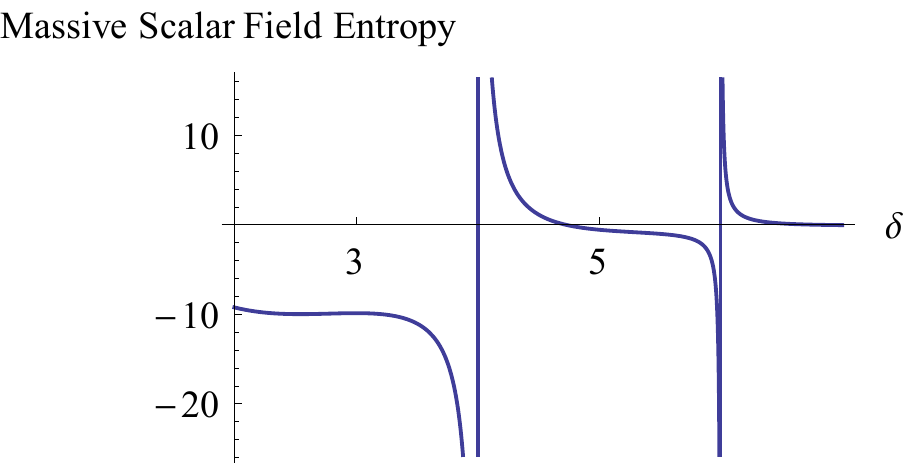}}
\\          
{\it Figure 3:  Dependence of massive scalar field entropy on $\Delta$  ($m=\omega=1$).}
\\
\\
Note that the divergence is out off the range in our studies because that the back-reaction about the BTZ space is small in prior assumption. In fact, in the unit $m=1$  the scalar field mass $M$ shall not larger then $1$ otherwise the back-reaction of scalar field on black hole will  be very strong and BTZ spacetime will be dramatically changed.  Therefore, while we have found the exact formula  it is in fact can be applied only near $\Delta=2$, i.e. $M<m$.
%%%%%%%%%%%%%%%%%%%%%%%
%%%%%%%%%%%%%%%%%%%%%%%
\subsection {Zero-Mode of Massive Real Scalar Field }
We first consider the zero-mode of massless scalar field
\be
\Phi= f(u)
\ee 
The massless field equation has solution
\be
f(u)=C_1+C_2\Big[\log(u^2 + m^2)-\log(u^2)\Big]
\ee
which, however is divergent on horizon $u=0$.  Thus we have not physical  solution to provide entropy.  

Let us turn to the massive case in which the normalized solution is
\be
f(u)=m^{-1 + \sqrt{1 + M^2}}~{\Gamma[{1\over2}(1 + \sqrt{1 + M^2})]^2 \over \Gamma[\sqrt{1 + M^2}]}~  \, _2F_1\Big[{1\over2}(1 - \sqrt{1 + M^2}~),{1\over2}(1 +\sqrt{1 + M^2}~), 1, -{u^2\over m^2}\Big]\nn\\
\ee
which becomes
\be
f(u)&=&m^{-1 + \sqrt{1 + M^2}}~{\Gamma[{1\over2}(1 + \sqrt{1 + M^2})]^2 \over \Gamma[\sqrt{1 + M^2}]}+{\cal O}(u^1)\\
f(u)&=&u^{-1+ \sqrt{1 + M^2}}\Big(1+{\cal O}(u^{-1})\Big)
\ee
Therefore, this solution is regular on horizon and has a  proper normalization likes as that required in  [1].

After the calculations the on-shell action is
\be
\log Z(\beta)=-\frac{2 \pi ^2 ~m^{2\sqrt{M^2+1}} \csc \Big(\pi 
   \sqrt{M^2+1}\Big) \sec \Big(\frac{1}{2} \pi  \sqrt{M^2+1}\Big)
   \Gamma \Big(\frac{1}{2}   \Big(\sqrt{M^2+1}+1\Big)\Big)}{\sqrt{M^2+1} \Gamma   \Big(\sqrt{M^2+1}\Big)^2 \Gamma   \Big(\frac{1}{2}-\frac{\sqrt{M^2+1}}{2}\Big)^3}\nn\\
\ee
and associated entropy is
\be
S_{\Phi}(M,m,0,\beta)&=&\frac{16 \pi ^4 ~m^{2\sqrt{M^2+1}} \sin    \Big(\frac{1}{2} \pi  \sqrt{M^2+1}\Big) \csc ^2\Big(\pi    \sqrt{M^2+1}\Big) \Gamma \Big(\frac{1}{2}    \Big(\sqrt{M^2+1}+1\Big)\Big)}{m \Gamma
   \Big(\sqrt{M^2+1}\Big)^2 \Gamma    \Big(\frac{1}{2}-\frac{\sqrt{M^2+1}}{2}\Big)^3}\nn\\
\ee
Now, let us discuss the properties of  $S_{\Phi}(M,m,0,\beta)$.

(1) In the case of  small $M$ 
\be
S_{\Phi}(M,m,0,\beta)&=&-m \pi^2 M^2+{\cal O}(M^3)
\ee
which means that there is not entropy in the limit of massless, which consistent with previous result.  

(2) As it appears $\csc^2\Big(\pi\sqrt{M^2+1}\Big)$ we see that the entropy will be divergent at some values of $M$,  as shown in figure 4.
\\
\\
\\
\scalebox{1}{\hspace{3cm}\includegraphics{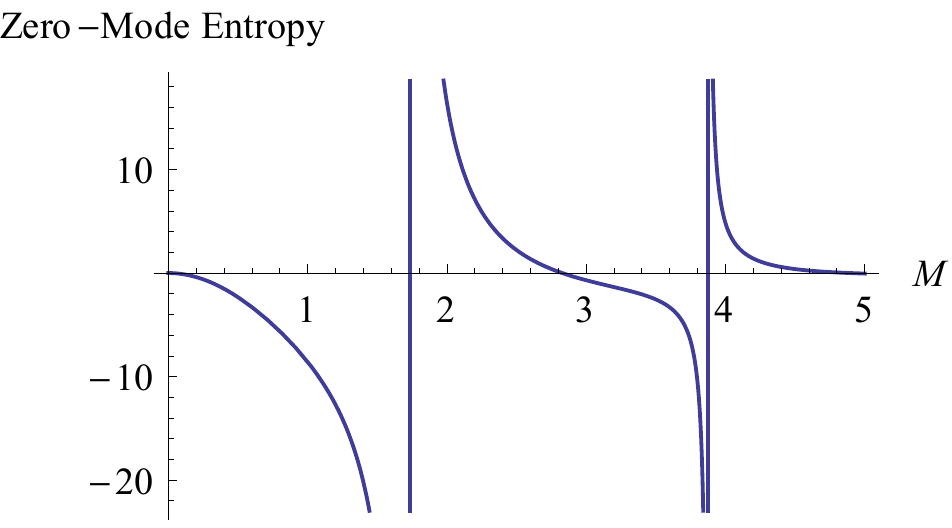}}
\\          
{\it Figure 4:  Dependence of zero-mode of massive scalar field entropy on $M$  ($m=1$).}
\\
\\
Note that, as  before,  the divergence is out off the range in our studies and  while we have found the exact formula  it is in fact can  be applied only if $M<m$.
%%%%%%%%%%%%%%%%%%%%
%%%%%%%%%%%%%%%%%%%%%%%
%%%%%%%%%%%%%%%%%%%%%%%
%%%%%%%%%%%%%%%%%%%%%%%
\section {DBI on BTZ Spacetime }
\subsection {Classical Solution of DBI on BTZ spacetime and Entropy}
We now study the back-reaction effect on BTZ black hole by D branes, which is described by DBI action [15].  We consider the case in coordinate $(t,u,\phi)$ with simplest  EM field 
\be 
A_\mu =(A_t(u),0,0)
\ee
Then, DBI action becomes  
\be
A&=&\int d^3x~\sqrt g~\Big[{1\over \pi^2\alpha'^2} \Big(\sqrt{g_{\mu\nu} +2\pi\alpha' F_{\mu\nu}}-\sqrt {g_{\mu\nu}} \Big)\Big]\\
&=&{1\over \pi^2\alpha'^2}~\pi\beta \int_0^\infty du\Big(-u + \sqrt{u^2 + 4\pi^2 \alpha'^2 (m^2 + u^2)  A_t'(u)^2}\Big)
\ee
and the associated conjugate momentum $\Pi$ is an integration constant which has a simple relation
\be
\Pi={1\over \pi^2\alpha'^2}{4\pi^2 (m^2 + u^2) \alpha'^2 A_t'(u)\over{\sqrt{u^2 + 4\pi^2\alpha'^2  (m^2 + u^2) A_t'(u)^2}}}
\ee
It gives solution
\be
A_t(u)=c_1+\frac{\Pi \log \left(4\alpha'\left(\sqrt{\pi ^2 \alpha'^2
   \left(m^2+u^2\right)-\Pi^2}+2\pi  \alpha' \sqrt{m^2+u^2}\right)\right)}{4\pi ^2 \alpha'^2}
\ee
where $c_1$ is another  integration constant. As before, by requiring that the solution $A_\phi(u)$ is zero on horizon and becomes $A_t(\infty)=\log (u)$  it is found that
\be
\Pi=4\pi^2 \alpha^2
\ee
Substituting it into the Lagrangian the on-shell gravitation action  is
\be
\log Z(\beta)_{DBI}&=&{1\over \pi^2\alpha'^2}\Big[{m\over 4}\Big(m - \sqrt{m^2 -4 \pi^2 \alpha'^2}\Big)-\pi^2 \alpha'^2   \log\Big({m + \sqrt{m^2 -4 \pi^2 \alpha'^2}\over 2u}\Big)\Big]
\ee
While there are divergent terms  ($r\rightarrow \infty$), which should be subtracted [1], they will do not contribute to the entropy and the associated black hole entropy  becomes 
\be
S_{DBI}(\beta)&=&-\partial_\beta\log Z(\beta)_{DBI}+\log Z(\beta)_{DBI}\nn\\
&=&{2m\over\alpha'^2} \left(-2 + 2 m\sqrt{1\over m^2 -4 \pi^2 \alpha'^2}~\right)\nn\\
&\approx&\frac{\pi^2}{m}+\frac{3\pi^4 \alpha'^2}{ m^3}+\frac{10 \pi^6 \alpha'^4}{m^5}+{\cal O}(\alpha'^6)
\ee
We thus find that the leading term is just the zero-mode entropy found in previous section of Maxwell field theory, the property that the leading order of DBI action is just the Maxwell field.  Note that the correction entropy form DBI is positive and is a decreasing function with respective to $m$, like as that in Maxwell case.  However, if $m \rightarrow \pi \alpha'$  then the correction  is divergent, as show in figure 5. The divergence shall not be shown in our studies because that the back-reaction about the BTZ space is small in prior assumption.
\\
\\
\\
\scalebox{1}{\hspace{3cm}\includegraphics{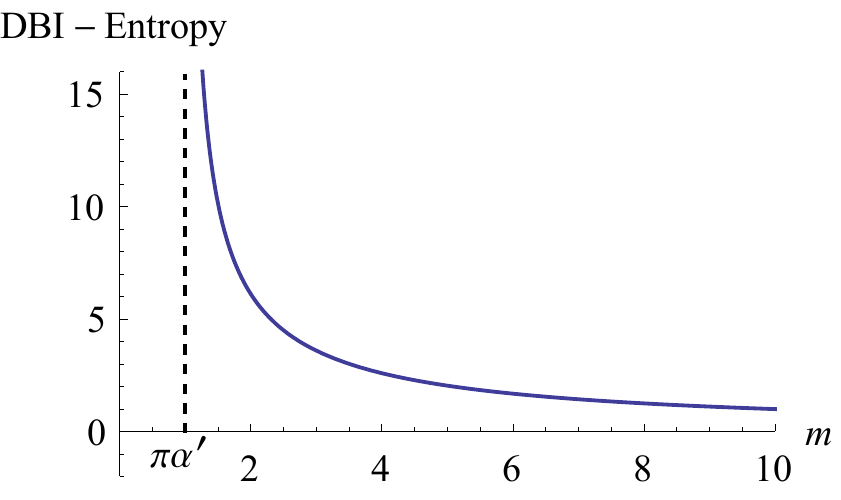}}
\\ 
\\         
{\it Figure 5:  Dependence of DBI entropy on $m$ in unit of  $2\pi \alpha'=1$.}
\\
%%%%%%%%%%%%%%%%%%%%%%
%%%%%%%%%%%%%%%%%%%%%%
%%%%%%%%%%%%%%%%%%%%%%%
\subsection {Classical Solution of DBI on BTZ spacetime and Area Law}
In this section we follow [1] to explicitly check that the classical solution which gives  the  black hole entropy precisely corrects the black hole area law.  While the original paper [1] checks the case of scalar field theory and [12]  checks the case of  Maxwell field, we will  investigate  the backreaction of the DBI field on the BTZ metric.  

The action is 
\be
-A=\int d^3x~\sqrt g~\Big[R-\Lambda -{1\over \pi^2\alpha'^2} \Big(\sqrt{g_{\mu\nu} +2\pi\alpha' F_{\mu\nu}}-\sqrt {g_{\mu\nu}} \Big)\Big]
\ee
with $\Lambda=2$. The Einstein equation  is
\be
G_{\mu\nu}\equiv R_{\mu\nu}-{g_{\mu\nu}\over 2}\Big(R-2\Big)=8\pi G~T_{\mu\nu}
\ee
Consider the case  in coordinate $(t,u,\phi)$.  After the ansatz that $A_\mu=(A_t(u),0,0)$ the stress tensors become
\be
T^u_u&=&{u^2\over 2\sqrt{u^2+4 \pi^2\alpha'^2A'_t(u)^2(u^2+m^2)}}\\
T^\phi_\phi&=&{1\over2}\sqrt{u^2+4 \pi^2\alpha'^2A'_t(u)^2(u^2+m^2)}
\ee
To proceed we see that the backreaction of DBI field will  modify the  BTZ metric, which can be written as [1]
\be
ds^2 =u^2dt^2+{du^2\over u^2+m^2+g(u)}+(u^2+m^2)(1+v(u))d\phi^2
\ee 
To the leading of perturbation (small $g(u),v(u)$)
\be
G_u^u&=&{g'(u)\over  2u}\\
G_\phi^\phi&=&{g(u)\over u^2+m^2}+{(u^2+m^2)v'(u)\over 2u}
\ee
Now, using the Einstein equations we can find that 
\be
v'(u)&=&\Big({g(u)\over u^2+m^2}\Big)'+8\pi G ~{u~\partial_uA'_t(u))^2\over \sqrt{u^2+4 \pi^2\alpha'^2A'_t(u)^2(u^2+m^2)}}
\ee
To proceed we can use following  two properties to perform the integration to find  the function $v(u)$ : 

(i) Using the property that $g(0) = 0$ due to the regularity condition for the metric at the origin and $g(r)/r^2\rightarrow 0$ at infinity we can get ride of  $g(u)$ (which is a total derivative) in integration [1].

(ii) We had derived in section 5.1 that 
\be
\Pi&=&4\pi^2 \alpha'^2=\frac{4 \pi ^2 \alpha'^2 \left(m^2+u^2\right)At'(u)}{\sqrt{u^2+4 \pi ^2 \alpha'^2
   \left(m^2+u^2\right)At'(u)^2}}\\
A_t'(u)&=&\frac{2\pi \alpha'^2 u}{\sqrt{\alpha'^4 \left(m^2+u^2\right) \left(u^2-4 \pi ^2 \alpha'^2+m^2\right)}}
\ee
Therefore
\be
v(0)&=&4\pi G~\int_{0}^\infty du~ \frac{4 \pi ^2 \alpha ^2 u^2}{\left(m^2+u^2\right) \sqrt{ \left(m^2+u^2\right) \left(\left(m^2+u^2\right)-4 \pi^2 \alpha^2\right)}}\\
&=&4\pi G~\frac{-4 \pi ^2 \alpha ^2 K\left(1-\frac{4 \pi ^2 \alpha ^2}{m^2}\right)+m^2 E\left(1-\frac{4 \pi ^2\alpha ^2}{m^2}\right)-m \sqrt{m^2-4 \pi ^2 \alpha ^2} E\left(\frac{m^2}{m^2-4 \pi ^2 \alpha
   ^2}\right)}{\alpha ^2 m}\nn\\
\ee
where $E$ and $K$ are the Elliptic functions. Using the black hole area law formula 
\be
S_{DBI}&=&{1\over 4G} {\cal A_H}={1\over 4G} \cdot 2\pi \cdot m\cdot  \Big(1+v(0)\Big)=S^{(0)}_H+\delta S_H
\ee
we finally find that 
\be
\delta S_H ={1\over 4G} \cdot 2\pi m ~v(0)\approx\frac{\pi^2}{m}+\frac{\pi^4 \alpha'^2}{2 m^3}+\frac{3 \pi^6 \alpha'^4}{4 m^5}+{\cal O}(\alpha'^6)
\ee
Thus the leading order expansion of $\delta S_T$  precisely gives the exact value calculated in previous section.  However, the higher orders do not match.  This is because that in our approach the backreaction of DBI field shall be weak [1] and perturbation of metric $v(u)$ is samll.  Thus, the result is reliable only in the case of small gravitation entropy, which is the case of leading order .

%%%%%%%%%%%%%%%%%%%%%%
%%%%%%%%%%%%%%%%%%%%%%
%%%%%%%%%%%%%%%%%%%%%%%
%%%%%%%%%%%%%%%%%%%%%%%
\section{Conclusions}
In this paper we extend the calculations of  the generalized gravitational entropy proposed in recent by  Lewkowycz  and Maldacena [1].  We study the back-reaction of massive classical scalar field, Maxwell field and DBI action on BTZ black hole entropy. We first follow [1] to prove that the classical solution of Einstein equation gives  the correction of horizon area in general black hole spacetime. We investigate the Maxwell  field in Coulomb gauge and Lorentz gauge on the BTZ spacetime.  We solve the classical solutions exactly and then use them to calculate the entropy which corresponds to the correction of area law.  The results is gauge independent.  We also study the massive scalar field and DBI action effect on the BTZ black hole. The back-reaction properties of these field on the BTZ black are found exactly. We explicitly check that the classical solution which gives  the  black hole entropy precisely corrects the black hole area law.  The dependences of mass and mode number on the  black hole entropy are then illustrated.

Our investigation can be summarized in below : 

1. Although the coordinates $(t,r,\theta)$ and $(t,u,\theta)$ are related by a simple transformation we find that, in some cases, while it is difficult to find the solution in one coordinate we can easily to find the solution in another coordinate.

2. The gravitation entropies of  scalar fields and Maxwell fields are  always negative value.  However, for the zero mode,  the entropy of  Maxwell field is positive while that of scalar field is negative.  We have no ideal to explain this property.

3. We have calculated  the exact values of  gravitation entropies from DBI action and  the backreaction of DBI field on the of black hole area law.  Using the fact that the  backreaction is  meaningful only on the weak field approximation we compare both values and  see that they are consistent in the leading order.

4. In contrast to the previous investigations [1,12], in this paper we let the mode number to be the arbitrary parameters.  After studying the the dependences of mass and mode number on the  black hole entropy we   find that the gravitational entropy could be divergent at  some values of mode numbers.  This means that the associated solution could dramatically modify the BTZ spacetime.  In this case the backreaction is vary large  and we shall not consider the problem in  the fixed BTZ spacetime, which is beyond the scope of  this paper.   

It remains to see whether these properties also show in other black hole spacetime.  Also, as quantum correction of black hole has universal property which is specified by the central charge [3],  it is interesting to see whether the classical correction has any universal property.  The problem is worthy to study. 
\\
\\
\\
\begin{center} {\bf REFERENCES}\end{center}
%%%%%%%%%%%%%%%%%%%%%%
%%%%%%%%%%%%%%%%%%%%%%
\begin{enumerate}
\item  A. Lewkowycz and J. Maldacena, ``Generalized gravitational entropy," JHEP 08 (2013) 090  [arXiv:1304.4926 [hep-th]].
\item  G. W. Gibbons  and S. W. Hawking, ``Action Integrals and Partition Functions in Quantum Gravity'' , Phys. Rev., D15 (1977) 2752.
\item S. N. Solodukhin, ``Entanglement entropy of black holes '', Living Rev. Relativity 14, (2011), 8 [arXiv:1104.3712 [hep-th]].
\item M. Srednicki ``Entropy and area'',  Phys. Rev. Lett., 71 (1993) 666 
[arXiv:hep-th/9303048].
\item  C.G. Callan and F. Wilczek, ``On geometric entropy", Phys. Lett., B333  (1994) 55 [arXiv:hep-th/9401072].
\item  V.P Frolov and I. Novikov, ``Dynamical origin of the entropy of a black hole", Phys. Rev., D48 (1993) 4545 [arXiv:gr-qc/9309001].
\item L.  Susskind, ``Some speculations about black hole entropy in string theory'', (1993) [arXiv:hep-th/9309145 [hep-th]].
\item F. Larsen and F. Wilczek, ``Renormalization of black hole entropy and of the gravitational coupling constant", Nucl. Phys., B458 (1996) 249 [arXiv:hep-th/9506066].
\item S. N. Solodukhin, ``Black hole entropy: statistical mechanics agrees thermodynamics'',  Phys. Rev., D54 (1996) 3900 [arXiv:hep-th/9601154].
\item N. D. Birrell and P. C. W. Davies,``Quantum Fields in Curved Space'', (Cambridge University Press, New York, 1982).
\item  L. Parker and D. Toms ,``Quantum Field Theory in Curved Spacetime'', (Cambridge University Press, New York, 2009). 
\item  Wung-Hong Huang, ``Generalized Gravitational Entropy of  Interacting Scalar Field and Maxwell Field'' , Physics Letters B. 739 (2014) 365[arXiv:1409.4893 [hep-th]].
\item  M. Banados, C. Teitelboim  and J. Zanelli, `` The Black hole in three dimensional space-time'',  Phys.Rev.Lett., 69 (1992)1849, [arXiv:hep-th/9204099].
\item  M. Banados, M. Henneaux, C. Teitelboim  and J. Zanelli, ``Geometry of the 2+1 Black Hole'', Phys.Rev.D48 (1993) 1506 [arXiv:gr-qc/9302012].
\item  C. G. Callan and J. M. Maldacena, ``Brane Dynamics From the Born-Infeld Action", Nucl.Phys. B513 (1998) 198 [arXiv:hep-th/9708147]
\item S. Carlip, ``The (2+1)-Dimensional Black Hole", Class.Quant.Grav.12 (1995) 2853 [arXiv:gr-qc/9506079]
\item R.B. Mann  and S.N. Solodukhin, ``Quantum scalar field on three-dimensional (BTZ) black hole instanton: Heat kernel, effective action and thermodynamics", Phys. Rev. D55 (1997) 3622 [arXiv:hep-th/9609085].
\item D.V. Singh and  S.  Siwach, ``Scalar Fields in BTZ Black Hole Spacetime and Entanglement Entropy",  Class. Quantum Grav. 30 (2013)235034[ arXiv:1106.1005  [hep-th]]
\end{enumerate}
\end{document}